\documentclass[preprint,12pt]{elsarticle}
\usepackage{amssymb}
\journal{Phsycia B: Condense matter}

\begin{document}

\begin{frontmatter}


\title {Rotating Bose-Einstein condensate in an square optical lattice: vortex configuration for ground state in Josephson junction arrays regime}

\author{Y. Azizi and A. Valizadeh}

\address{ Institute for Advanced Studies in Basic Sciences,
Zanjan 45195-1159, Iran}

\ead{azizi@iasbs.ac.ir; valizade@iasbs.ac.ir}



\begin{abstract}
We consider a rotating Bose-Einstein condensate in a square
optical lattice in the regime in which the Hamiltonian of the
system can be mapped onto a Josephson junction array. In an
approximate scheme where the couplings are assumed uniform, the
ground state energy is formulated in terms of the vortex
configuration. The results are compared with experimental results and also previously
reported results for frustrated XY model. We also show that vortex configuration is robust with
respect to change of couplings and therefore the
results remain valid when we consider more realistic model with
non-uniform couplings.
\end{abstract}

\end{frontmatter}

\section{Introduction}
After the experimental realization of Bose-Einstein condensation
(BEC)\cite{R1}, it becomes an interesting topic in
science\cite{R2,R3,R4,R5}. This physical system has close relation
with other important physical problems in condense matter, e.g.
Bloch Oscillations\cite{R6,R7,R8,R9}, Wannier-Stark
Ladders\cite{R10}, Josephson junction
arrays\cite{R11,R12,R13,R14,R15,R16,R17} and superfluid to the
Mott-insulator transition\cite{R18,R19,R20,R21,R22}.

Behavior of a rotating BEC in an optical lattice is similar to
that of a Superconductor in the magnetic field
\cite{R23,R24,R25}. When there is just trap potential without the
lattice structure, the Abrikasov vortex lattice can be observed
just like type II Superconductors\cite{R11,R12,R14,R15,R26,R27,R28,R29,R30,R31}.
When the lattice structure is added to system, the question of the
vortex lattice in ground state becomes more complicated\cite{R32}. 
There are many experimental studies on
this system with optical lattices of different symmetries, specially with square lattice\cite{R30a,R30b,R30c}. 
With some criteria this problem can be mapped to problem Josephson Junction arrays
(JJAs) and numerical experiments show that vortex lattice is same for both the systems\cite{R12,R14}.

We focus on study of problem in this regime and extend our
previous results\cite{R33} for an optical lattice with square
symmetry. We calculate vortex configuration for few rational 
rotation numbers and show that there is good agreement with
experimental results on rotating BEC and reported result for
Josephson junctions lattices\cite{R43}.

Structure of paper is as follows: In section II we formulate
problem in JJA regime. In section III we focus on vortex lattice
with rational rotation numbers and compare our result with previous experimental
and numerical results. In the last section main results are highlighted and suggestions for 
future works have been given.

\section{Rotating BEC in an optical lattice}
Hamiltonian for a BEC in an optical lattice with a rotation
frequency $\Omega \hat{z}$ is,
\begin{equation}\label{bechami}
\hat{H}=\int dr \hat{\Psi}^\dagger[\frac{(-i\hbar \nabla - M\Omega
\hat{z}\times
r)}{2M}+V_{ext}+\frac{g}{2}\hat{\Psi}^\dag\hat{\Psi}-\mu
]\hat{\Psi}
\end{equation}
where $M$ is the atomic mass and $g=4\pi \hbar^2a/M$ the coupling
constant with the $s$-wave scattering length $a$. Conservation of
the total number of particles is ensured by chemical potential
$\mu$. The external potential $V_{ext}=V_h+V_O$ consists of two
parts: modified harmonic potential
$V_h=M(\omega_{\perp}^2-\Omega^2)r^2/2+M\omega_z^2z^2/2$ and the
potential of the optical lattice $V_O$, which may be chosen as
periodic\cite{R34}, quasiperiodic\cite{R34,R35}, or
random\cite{R41}. For example, the potential with square symmetry
can be written as $V_O=V_0[sin^2(kx)+sin^2(ky)]$, with the
periodicity $\pi/k$.

For a large $\omega_z$, we can suppose that the system is frozen
in the axial direction. If the energy due to the interaction and
the rotation is small compared to the energy separation between
the lowest and the first excited band, the particles are confined
to the lowest Wannier orbitals\cite{R14}. Therefore We can write
$\psi$ in the Wannier basis as
\begin{equation}\label{wannier}
\hat{\Psi}(r)=\sum_{i=1}^N\hat{a}_iw_i(r)exp(\frac{im}{\hbar}\int_{r_i}^r
A(r').dr'),
\end{equation}
where $A=\Omega\hat{z}\times r$ is the analog of the magnetic
vector potential, $w_j(r)=w(r-r_i)$ the normalized Wannier
wave-function localized in the $i$-th well and
$\hat{m}_i=\hat{a}_i^{\dag}\hat{a}_i$ the number operator.
Substituting \ref{wannier} in the Hamiltonian \ref{bechami} leads to
the Bose-Hubbard model in the rotating frame:

\begin{eqnarray}\label{bosehubbard}
\hat{H}&=&-t\sum_{<i,j>}(\hat{a}^{\dag}_i\hat{a}_je^{-A_{ij}}+h.c.)\nonumber\\
&+&\sum_i(\epsilon_i-\mu)\hat{m}_i+\frac{U}{2}\sum_i\hat{m}_i(\hat{m}_i-1),
\end{eqnarray}
where $\langle i,j\rangle$ means $i$ and $j$ are nearest
neighbors, $t\approx -\int dr w_{j}^*(-\hbar^2\nabla^2/2m+V_O)w_j
$ is the hopping matrix element, $\epsilon_i=\int dr w^*_iV_hw_i$
an energy offset of each lattice site and $U=g\int dr|w_i|^4$ the
on-site energy. The rotation effect is described by $A_{ij}$
proportional to the line integral of $A$ between the $i$-th and
$j$-th sites: $A_{ij}=(m/\hbar)\int_{r_i}^{r_{j}}A(r').dr'$.

\begin{figure}[ht!]
\centerline{\includegraphics[width=15cm]{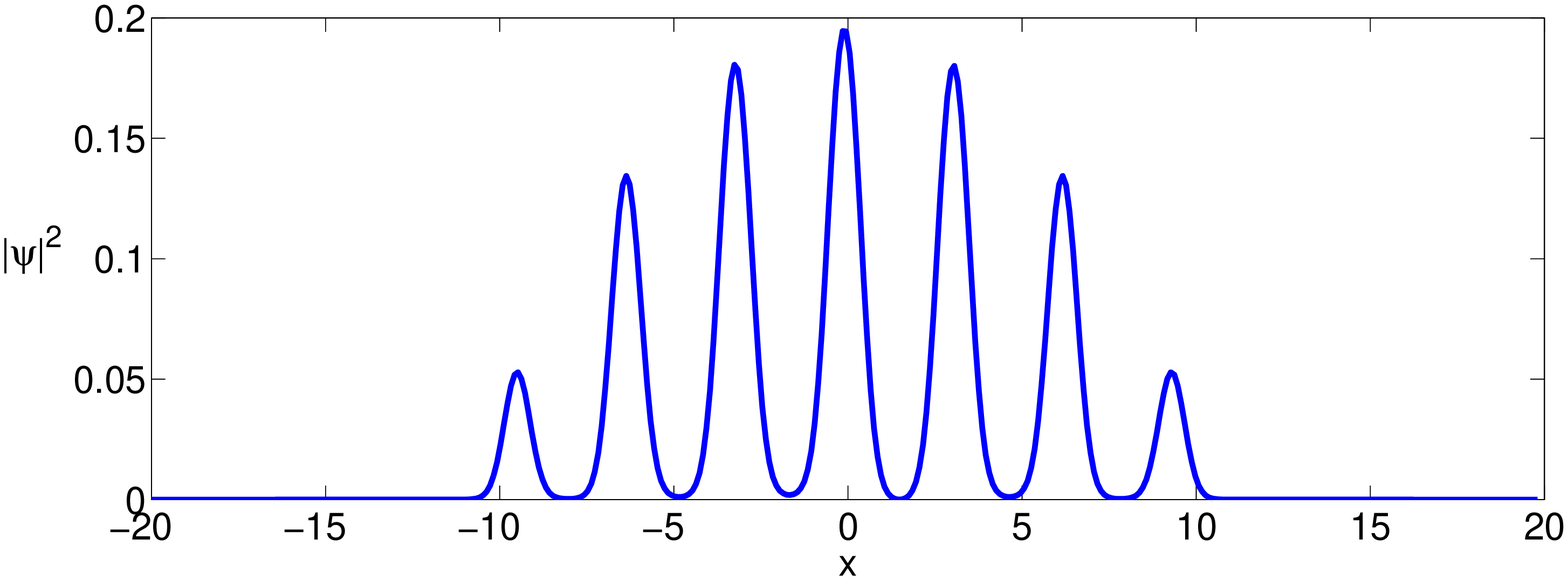}}
\vspace{0.1cm} \caption{Solution of GP equation for a periodic
one dimensional potential. It can be seen that $|\Psi|^2$ is
enveloped by an inverted parabola shape.} \vspace{-0.0cm}
\label{oned}
\end{figure}

If the number of atoms per site is large ($n_i\gg 1$), the
operator can be expressed in terms of its amplitude and phase,
and the amplitude by the $c$-number as $\hat{a_i}\approx
\sqrt{m_i}e^{i\hat{\theta}_i}$. Using
$\hat{m}_i=m_i-i\partial/\partial\theta_i$ and
$\hat{\theta}_i=\theta_{i}$, Eq. \ref{bosehubbard} reduces to the
quantum phase model:
\begin{equation}
\hat{H}=-\sum_{<i,j>}J_{ij}cos(\theta_i-\theta_{j}+A_{ij})-\frac{U}{2}\sum_i\frac{\partial^2}{\partial\theta_i^2},
\end{equation}
where $J_{ij}=2t\sqrt{m_im_{j}}$. Also, the atoms are assumed to
be distributed such as they satisfy the condition
$\epsilon_i+Un_i=\mu$. The magnitude of $J_{ij}$ decreases from
the central sites outward since $m_i$ has the profile of an
inverted parabola as we can see from solution of 1D case in Fig.
\ref{oned}.

Eq. 2.4 is just the Hamiltonian of a lattice of small Josephson
junctions with inhomogeneous coupling constants $J_{ij}$. With
$J_{ij}\gg U$, we can neglect the kinetic energy term; Moreover,
we suppose that coupling constants are equal $J_{ij}=J$. We also
check the robustness of the results for non-uniform coupling.

This regime can be achieved when the depth of the optical lattice
$V_0$ is between 18 and 25$E_R$ ($E_R=\hbar^2k^2/2M$ is the recoil
energy) and the average particle number is $\bar{m}\approx
170$\cite{R42}. Then we have:
\begin{equation}\label{hamiltonian}
H=-\sum_{<i,j>}cos(\gamma_{ij}),
\end{equation}
with $\gamma_{ij}=\theta_{i}-\theta_{j}-A_{ij}$. Our purpose is
to find the minimum of this Hamiltonian as a function of the
condensate phases $\theta_{i}$ with two constrains as follows.
First constraint originates in the single valuedness of the wave
function which leads to the quantization of vorticity:
\begin{equation}\label{fq}
\sum \gamma_{ij}=2\pi(n_k-a_kf)
\end{equation}
where sum is over $k$th plaquette, $a_k$ is the area of $k$th
plaquette, and $n_k$ is integer. In the case of a condensate with
the rigid-body rotation, {\it the frustration parameter} is
defined as $f=2\Omega/\kappa$ with the quantum circulation
$\kappa=h/m$. In two dimensional arrays, frustration represents
density of vortices per the number of the plaquettes \cite{R43}.

The second constraint comes from conservation of the particles
numbers. The particle current from the nodes can be assumed as
the derivation of the energy function with respect to the
corresponding phase\cite{R44}. This leads to
\begin{equation}\label{ccons}
\sum sin(\gamma_{ij})=0,
\end{equation}
where sum is over all $j$ which is connected to $i$th node.

Since Hamiltonian is periodic, $n_k$ can be $\lfloor a_kf\rfloor$
or $\lfloor a_kf\rfloor+1$, and we can replace them by $n_k$
which is zero or one, and replace $a_kf$ with $a_kf-\lfloor
a_kf\rfloor$. These two set of equations can completely determine
all $\gamma_{ij}$ for a given set of $n_k$. Also, for a lattice,
number of plaquettes plus number of nodes is equal to number of
connections minus one\footnote{This is Euler formula for polygon
lattice excluding exterior face.}; therefore one of the equations
(e.g. one of particle conservation equations) can be neglected.
Assuming the particle flow from the nodes are small, we can
linearize the Eq.\ref{ccons} i.e. $sin{\gamma_{ij}\approx
\gamma_{ij}}$. Now, we have a linear set of equations which must
be solved for a given set of integers $n_k$, i.e. for a vortex
configuration. Energy of this vortex configuration is denoted by
$E_{\{n\}}$. Then instead of original minimization scheme, we can
use the set of these energies for minimization of Hamiltonian. It
means that minimum of Hamiltonian correspond to a vortex
configuration which minimizes these set of energies, with number
of vortex configuration equal to $2^{N}$ where $N$ is number of
plaquettes in the lattice.

\section{vortex configuration in ground state: checkerboard structure}
We consider the square lattice and present the result of above
formulation for some rational values of $f$. When the number of the
plaquettes becomes large, the checkerboard vortex structure of
ground state can be observed\cite{R43}. The same pattern can be
observed in the trapped BEC\cite{R30a,R30b,R30c}. We show that our
formulation has the same result.

\begin{figure}[ht!]
\centerline{\includegraphics[width=10cm]{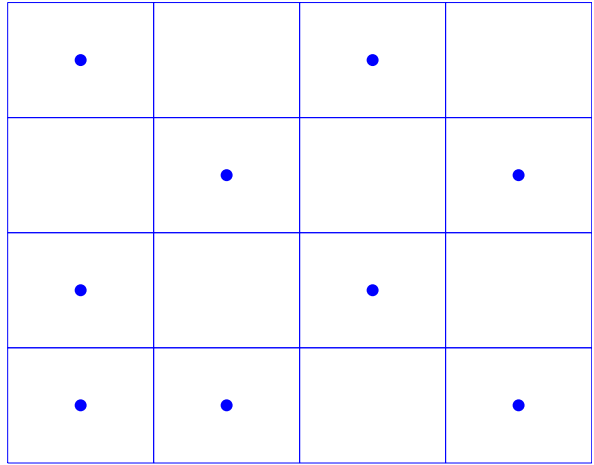}}
\vspace{0.1cm}
 \caption{Minimum vortex configuration for $4\times 4$ square lattice for $f=1/2$}
\vspace{0.5cm} \label{sq2d44}
\end{figure}

\begin{figure}[ht!]
\centerline{\includegraphics[width=10cm]{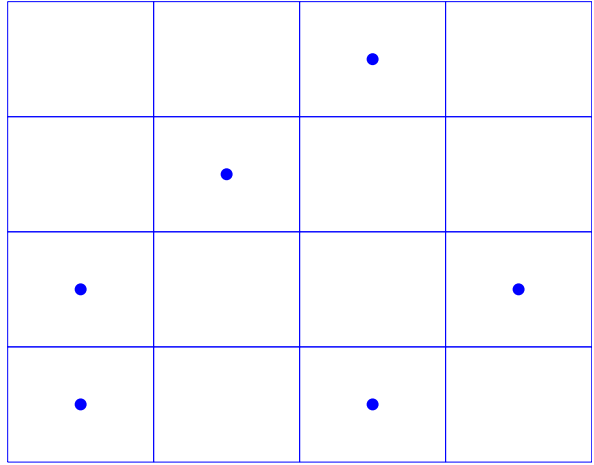}}
\vspace{0.1cm}
 \caption{Minimum vortex configuration for $4\times 4$ square lattice for $f=1/3$}
\vspace{0.5cm} \label{sq2d442}
\end{figure}

Set of the equations \ref{fq} and the linearized version of
\ref{ccons} for the square lattice can be written as
\begin{eqnarray}
C\gamma=b,
\end{eqnarray}
where $\gamma$ is a the vector representation of the gauge
invariant phase differences and $C$ is the matrix of coefficients:,
\begin{eqnarray}
C_{i,k}&=&-1\\
C_{i,k-1}&=&1\\
C_{i,k-m}&=&1\\
C_{i,k-m-1}&=&-1,
\end{eqnarray}
for $i=1,..,nm-1$ where $k=\lfloor i/m\rfloor+i$, $n$ is number of
rows and $m$ is number of columns; and for $i=nm,...,2mn-m-n$,
\begin{eqnarray}
C_{i,k}&=&1\\
C_{i,k+m}&=&1\\
C_{i,k+m-1}&=&-1\\
C_{i,k-m}&=&1,
\end{eqnarray}
where $k=(m-1)\lfloor (i-mn+1)/m\rfloor$. $b$ is zero for
$i=1,...,nm-1$ and
\begin{eqnarray}
b_i=2\pi (n(i-mn+1)-a(i-mn+1)))
\end{eqnarray}
for $i=mn,...,2mn-m-n$. These equations can solved for any given
$m,n$ and set of $n_i$ and $a_i$. The result for the minimum
vortex configuration for an square lattice with $a_i=1$, $m=n=4$
is shown in Fig. \ref{sq2d44} for $f=1/2$. In this case for $0\le
f \le 1/2$, $n_i$ could be zero or one. The plaquettes with points
in Fig. \ref{sq2d44} are those with $n_i=1$. The minimum vortex
configuration shows the checkerboard pattern which is known for
frustrated $XY$ model\cite{R43}.
For $f=1/3$, the minimum vortex configuration, is shown in
Fig.\ref{sq2d442}. Again the checkerboard pattern is seen for
the vortex configuration in the ground state. In both cases, we
can see that there is an extra vortex in the boundary of the
lattice. This is a boundary effect and can be ignored in large lattice limit. These results are in agreement with
experimental results on rotating BEC\cite{R30a,R30b,R30c}.

When number of plaquettes increases, number of possible vortex
configurations, increases and we can use a combination of this
method with Monte-Carlo method\cite{R33}.

\section{Robustness of minimum vortex configuration with respect to coupling constants}
\begin{figure}[ht!]
\centerline{\includegraphics[width=10cm]{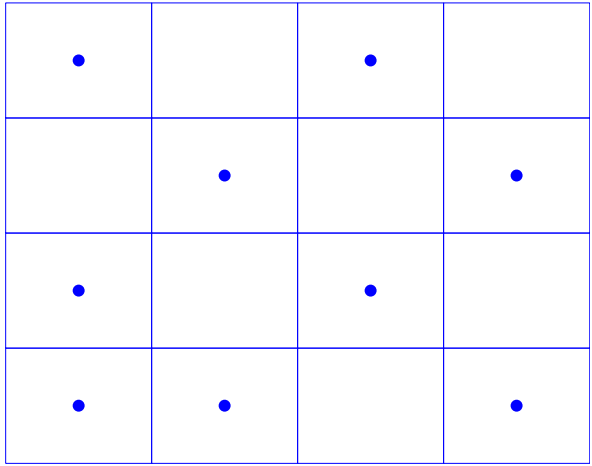}}
\vspace{0.1cm}
 \caption{Minimum vortex configuration for $4\times 4$ square lattice for $f=1/2$ with non-uniform coupling.}
\vspace{-0.5cm} \label{robust1}
\end{figure}

We see that in our approximation, the coupling of JJA changes as
the inverted parabola. Here, we discuss the effect of changing the coupling constants on the minimum vortex configuration. 
We consider an inverted parabola
$m_i$,
\begin{eqnarray}
m_{ij}=16ij(i-N_R-1)(j-N_C-1)/[(N_C+1)^2(N_R+1)^2]
\end{eqnarray}
for $ij$th site of square lattice, where $N_C$ is the number of
columns and $N_R$ is the number of rows. The coupling between two
nearest neighbor sites on the lattice can be found from
$J_{(i,j),(k,l)}=\sqrt{m_{ij}m_{kl}}$ which gives a better
approximation comparing to the uniform couplings. Result for the same situation as
Fig. \ref{sq2d44}, is shown in Fig. \ref{robust1}. It is seen that the 
minimum vortex configuration is not affected by the change of the
coupling configuration. We have also checked the results for random $m_{ij}$ and we found the 
same minimum vortex configuration.
\section{Conclusion}
We study a rotating BEC in a square optical lattice in a regime
which Hamiltonian of the system can be mapped onto JJA. In this
regime, we formulate ground state of system in terms vortex
configuration. Our results in uniform coupling case, show that
vortex configuration in ground state has checkerboard pattern. We have
also checked the non-uniform coupling in the form $DJD$, which $D$
is a diagonal matrix, and $J$ is original uniform coupling
matrix. The results for this case show that minimum vortex
configuration is robust with respect to the change of coupling
matrix.

\end{document}